\documentclass[runningheads]{witkowski}

\usepackage[utf8]{inputenc}
\usepackage[square,numbers]{natbib}

\usepackage{graphicx}
\usepackage{amsfonts}
\usepackage{amsmath}
\usepackage{comment}
\usepackage{underscore}
\usepackage{booktabs}
\usepackage{hyperref}
\usepackage{enumerate}
\usepackage{caption}
\usepackage{subcaption}
\usepackage{setspace}

\begin{document} 

\doublespacing

\title{
How to Make Swarms Open-Ended? \\Evolving Collective Intelligence Through a Constricted Exploration of Adjacent Possibles
}
\titlerunning{How to Make Swarms Open-Ended?} 

\author{
Olaf Witkowski\inst{1,2,*}
\and 
Takashi Ikegami\inst{3}
}

\authorrunning{O. Witkowski and T. Ikegami} 

\institute{Earth-Life Science Institute, Tokyo Institute of Technology, Tokyo, Japan \and
Institute for Advanced Study, Princeton, USA \and
University of Tokyo, Tokyo, Japan\\
* okw@elsi.jp
}

\maketitle

\begin{abstract} 
We propose an approach of open-ended evolution via the simulation of swarm dynamics. In nature, swarms possess remarkable properties, which allow many organisms, from swarming bacteria to ants and flocking birds, to form higher-order structures that enhance their behavior as a group. Swarm simulations highlight three important factors to create novelty and diversity: (a) communication generates combinatorial cooperative dynamics, (b) concurrency allows for separation of timescales, and (c) complexity and size increases push the system towards transitions in innovation. We illustrate these three components in a model computing the continuous evolution of a swarm of agents. The results, divided in three distinct applications, show how emergent structures are capable of filtering information through the bottleneck of their memory, to produce meaningful novelty and diversity within their simulated environment.

\end{abstract}

\section{OEE}

Life  has  been  evolving  on  our  planet  over  billions  of years,  undergoing  several  major  transitions which transformed the way it stored, processed and transmitted information. All these transitions, from multicellularity to the formation of eusocial systems and the development of complex brains, seem to lead to the idea that the evolution of living systems is \textit{open-ended}. In other words, life appears to be capable of increasing its complexity indefinitely 
Another formulation of open-endedness, echoed by Standish \cite{standish2003open} and Soros \cite{soros2014identifying}, is that open-endedness depends fundamentally on the continual production of novelty. 
Life keeps uncovering new inventions, in a process which never seems to stop.

Since the 1950's, open-ended evolution (OEE) has been a central topic of research for artificial life approaches to the fundamental principles of life. Soon after, John Von Neumann \cite{von1966theory} has contributed to the issue as well, with his early model of self-reproducing automata. 
Since 2015, a series of workshops have been taking place at Artificial Life conferences \citep{taylor2016open}, the last of which\footnote{at ALIFE 2018, in Tokyo} was a launchpad for the present special issue. 
In general, an evolutionary system is considered to be open-ended when it is able to endlessly generate diverse novel entities of growing complexity. 
Engineering open-ended systems in the lab is not easy, and the main obstacle is that the designed evolutionary systems are subject to a thermodynamic drift making them collapse into equilibrium states. Once local optima are reached, they do not produce novelty anymore, which bounds their complexity and diversity.

Innovation seems to emerge from collective intelligence, a phenomenon which refers to groups or networks of agents that develop together the ability to enhance the group's cognitive capacity or creativity. This is reminiscent of the ongoing innovative process of science, which does not have any other fixed objective than the production of new knowledge, but makes discoveries mostly through accidents. Stuart Kauffman advocated for the idea of the adjacent possible, claiming that a biosphere can be viewed as a secular or long-term trend and it can maximize the rate of exploration of the adjacent possible of an existing organization \citep{kauffman2000investigations, kauffman2003adjacent}. Ikegami et al. (2017) \cite{ikegami2017life} builds on that idea to explain how, in terms of evolutionary transitions \citep{MaynardSmithSzathmary1997}, a new stage (e.g. multicellular oranism) of evolution may be produced without any information being passed on from the previous stage (e.g. from single cells). Rather, structural properties are assembled, producing a stepping stone to the next level of innovation.

These structural properties of a collective group can be compared to a bottleneck that acts as a filter on several levels of the system, implementing computation that is not present in any of its parts. Part of this idea is analog to Tishby and Polani (1999) \cite{tishby1999information}, where the information is squeezed through a bottleneck, to extract relevant or meaningful information from an environment. 
The resulting ``filtered'' information through the bottleneck, retains only the features most relevant to general concepts.

In this paper, we present three ``C'' factors that we deem important for novelty and diversity. We then introduce a swarm model to study these factors, applied in three different studies. We conclude with a discussion on open-endedness at large, framing the three factors in terms of the emergence of collective intelligence in swarm simulations.

\section{Conditions for OEE}

The OEE literature has proposed various conditions that are supposed to lead to the successful production of OEE in a system. 
Number of studies have attempted to formalize necessary conditions for OEE \citep{holland1999echoing, conrad1970evolution}. Taking a recent example, Soros and Stanley (2014) \cite{soros2014identifying} suggest four conditions at the scale of single reproducing individuals in the system, which should each fulfill some nontrivial minimal criterion, be able to create novelty, act autonomously, and dispose of access to unbounded memory.

Such papers have typically been proposing their own model, to demonstrate the importance of the hypothesized conditions for the emergence of OEE within it. However, most evolutionary algorithms seem to either converge very quickly to a solution, or get stuck in a confined area of the search space. Either way, they don't seem to be able to intrinsically generate the amounts of complexity and novelty we find in nature, even at a scale.

Although this failure of simulated evolution to match open-ended properties found in natural evolution may be explained by shorter timescales, in principle one would have expected decades of efforts, and increasingly larger resources poured into research in evolutionary computation, to have unlocked more of its potential to create novelty. However, even the latest technologies don’t seem to keep their inventivity. In general, promising models \citep{lenski2003evolutionary, lehman2011evolving, goodfellow2014generative, greenbaum2016digital, silver2017mastering} that manage to demonstrate at least a few phases transitions or creative leaps -- not necessarily with evolutionary computation -- seem to have one common denominator of containing several structural bottlenecks which filter relevant information through them, as a catalyst of creativity.

What seems to be missing to achieve general OEE? We choose to emphasize three ``C'' candidates which we see as worth pursuing -- Communication, Concurrency, and Complexity:

\begin{enumerate}[(a)]
    \item Communication:  
    constricted information flows between parts of the systems allow for synergy and cooperation effects. 
    
    \item Concurrency: 
    the creation of separate space and time scales requires concurrent, nondeterministic, asynchronous models. 
    
    \item Complexity: mere system growth can boost novelty and diversity.
    
\end{enumerate}

These are the three C-factors on which we choose to concentrate, in this paper. Next, will expand on each of them a little further, before proposing how to apply them in concrete models.

We will now expand on these three points, before presenting concrete examples, with Study 1, 2 and 3.

\subsection*{Communication}

This first point addresses synergies and coordination between components of the systems, using information transfers. One well known example of OEE is combinatorial creativity in human language, where syntactical rules are capable of producing infinite well-formed structures using recursion, thus making the number of potential sentences unbounded \citep{hauser2002faculty}. Although these may seem slightly dated remarks, at the advent of language studies based on artificial life systems \citep{kirby2002natural}, it is promising to focus on the cultural layer of dynamics, that lives on top of the main layer of entities. For example, in the case of web services (social tagging networks), we can analyze how combinatioral complexity is effective in evoking OEE.
In cellular automata, one may want to study the interactions between gliders or other patterns. In artificial chemistry, one may want to look at information flows between types of molecules or replicators. In agent-based modeling, perhaps establishments of protocols between agents or groups of agents can become a factor to focus on.

Communication naturally adds relevant computing filters on unexploited information flows, effectively increasing the bandwidth of useful information flows within the system per clock cycle, communication offers a layer for metadynamics at a different timescale from the first-order dynamics. This induces a separation of timescales, thus doubling the system's capacity to implement learning mechanisms. Designing information to circulate between a sub-entities of the system forces the creation of more structural bottlenecks. 

We propose information exchanges as a central mechanism promoting open-ended evolution. From information flows in groups of individuals, a system can boost its own production of creativity to achieve indefinite complexity. Examples are detailed in Study 1 and 2, below.

\subsection*{Concurrency}

In many situations, a system cannot scale up to larger space-time scales as it is.  We need to add some ingredient to make it work in the larger scales. One such remedy is to give it asynchronous updates. Removing the global clock is needed to make larger systems function consistently without constantly checking local consistency. On the other hand, we know that cellular automata (CA) tend to lose their complexity by adopting asynchronous dynamics. Yet asynchrony is an original natural phenomena difficult to bring to artificial systems.

According to Dave Ackley \cite{ackley2015artificial}, models should be indefinitely scalable, ruling out deterministic, synchronous models (such as simple cellular automata), and suggesting nondeterministic, asynchronous ones. 
Bersini et al. (1994) \cite{bersini1994asynchrony} proposed that asynchronous rather than synchronous updating may be key factor in inducing stability in simulations. Although they were examining a variant of cellular automata, their results, based on an analysis of the Lyapunov exponent, indicated the responsibility of asynchrony for sensitivity of the update function. Ackley and Ackley (2015) \cite{ackley2015artificial} propose to use asynchrony.

The concurrency is also closely related to the ability a system has to evolve separate timescales. Although highly contingent on other properties of a simulation, the capacity to develop heterogeneous timescales often constitutes a barrier to producing intrinsic novelty. Researchers used to separate lifetime learning and evolutionary learning, as two distinct mechanisms \citep{nolfi1990learning}. However, the effects of accumulating and filtering information into and out of one system's memory occurs at a much more continuous range of different timescales. In nature, from phenotypic plasticity, to maternal effects, to sexual selection, or to gene flow, many events have their time scales intricately interlaced. We will address this in particular in Study 1 and 2.

\subsection*{Complexity}

We have no grounds to argue that nature is its own only possible realization, since there would be no satisfactory explanation for that. One important feature of nature though, is its complexity, which can translate into both system size and landscape complexity. In the simplest of all cases, complexity can be reached merely with large population sizes. Ikegami et al. (2017) \cite{ikegami2017life} proposed that large groups of individuals, given the right set of structural characteristics, may be the main driver for emergence. They discussed this hypothesis in relation to large-scale boid swarm simulations \citep{reynolds1987flocks}, in which the nucleation, organization and collapse dynamics were found to be more diverse in larger flocks than in smaller flocks.

Collective behaviors can be qualitatively different by increasing the number of agents, i.e. a colony or group size. In the actual observation, e.g., the individual bees change their behaviors depending on the colony size. Also the fish change their performance of sensing the environmental gradient depending on the school size. 
In previous works \citep{mototake2015simulation}, we simulated half a million birds flock using a boids model \citep{reynolds1987flocks,toner1998flocks} and found that qualitatively different behavior emerges when the total number of individuals exceeds a few thousands or so. Flocks of different sizes and different shapes interact to diminish some flocks but to generate new ones. Different types of fluctuation become dominant in different size of flocks. A correlation of the local density fluctuation becomes dominant in the larger size flocks and that of the velocity fluctuation dominates in the smaller size flocks. An example of that is offered in Study 3.

Stretching the argument on size, environmental complexity is definitely necessary to a certain extent to create complex behaviors. Only with richer environments, encompassing complex distributions of energy resources and ways for systems to survive, can emerging individuals explore a rich set of strategies and increasingly increasingly complex solutions. As mentioned earlier, evolutionary landscapes have become an important concept in biology to analyze the dynamics at play in an ecosystem. 

The picture to have is the one of a unit of selection (e.g. a gene, among many other options) being represented by a point in a multidimensional search space. That space is typically given as many dimensions as there are degrees of freedom for the entity to vary and evolve in the space (e.g. combination of nucleotide sequences). The search space is mapped onto an additional dimension, which is usually the reproductive success, or fitness. The shape of the fitness across all degrees of freedom of the system have a strong impact on the dynamics that it can achieve. Malan et al. (2013) \cite{malan2013survey} identifies eleven characteristics of fitness, which make them more or less difficult to solve. These characteristics include the degree of variable inter-dependency, noise, fitness distribution, fitness distribution in search space, modality, information to guide search and deception, global landscape structure, ruggedness, neutrality, symmetry, and searchability. 

In evolutionary systems, richer environments, benefiting of a complex distribution of energy sources and ways for systems to survive, give rise to richer sets of pathways. The larger the search spaces, the more complex fitness functions are potentially evolved. Another way is to make the environment a more complex function of time, which the agents will need to learn in order to extract more energy from it. In Study 1 and 2, we present results suggesting that simulations should be ensured to provide sufficient system complexity in terms of the environment of agents. 

Simulation time and memory, though not mentioned yet, are important components to consider. Computationally, the whole course of evolution on Earth is like a single run of a single algorithm that invented all of nature, and seems like it will never end. One obvious difference is the size of the systems, which might be the missing element to get ever-greater emergent complexity and novelty through very long time. However, we do not insist on this component in this paper, as we consider trivial that a system with too low computational power will not be able to achieve OEE to any extent.

Similarly, there is point to be made about endo-OEE, producing novelty from within, against exo-OEE, which makes use of input from outside the system. Picbreeder \citep{secretan2008picbreeder}, for example, explicitly requires external human input to function, which makes it a debatable generator of OEE. Nevertheless, OEE is not about new information, but rather inventions achieved by the system. In that respect, swarms are a promising model: without increasing ensemble size, they let us focus on how coordination patterns self-organize,  generating intrinsic novelty. To give another example, even increasing the number of neurons in a neural network still requires neurons to differentiate themselves, and create coordinated networks before they get to foster innovative ideas.

We exemplify the importance of size and complexity in Study 1 and 3, while discussing how to make simulations parallelizable, to save considerable amounts of time and memory by distributing them over many machines.

\section{Concurrent evolutionary neural boids model}

The model we choose to present puts together the abovementioned series of features, as a means to promote the system's open-endedness. We give some details here, and will go over the details of several applications of it in the next section. The evolutionary system is an agent-based simulation, based off Reynolds' boids model \citep{reynolds1987flocks}. 

The boids model was based on simple rules computed locally, allowing to simulate flocks of agents moving through artificial environments. As with most artificial life simulations, boids showed emergent behavior, that is, the complexity of boids arises from the interaction of individual agents adhering to a set of simple rules of separation (steer to avoid crowding local flockmates), alignment (steer towards the average heading of local flockmates, and cohesion (steer to move toward the center of mass of local flockmates).

\begin{figure}[ht!]
\begin{center}
\includegraphics[width = 0.75\textwidth]{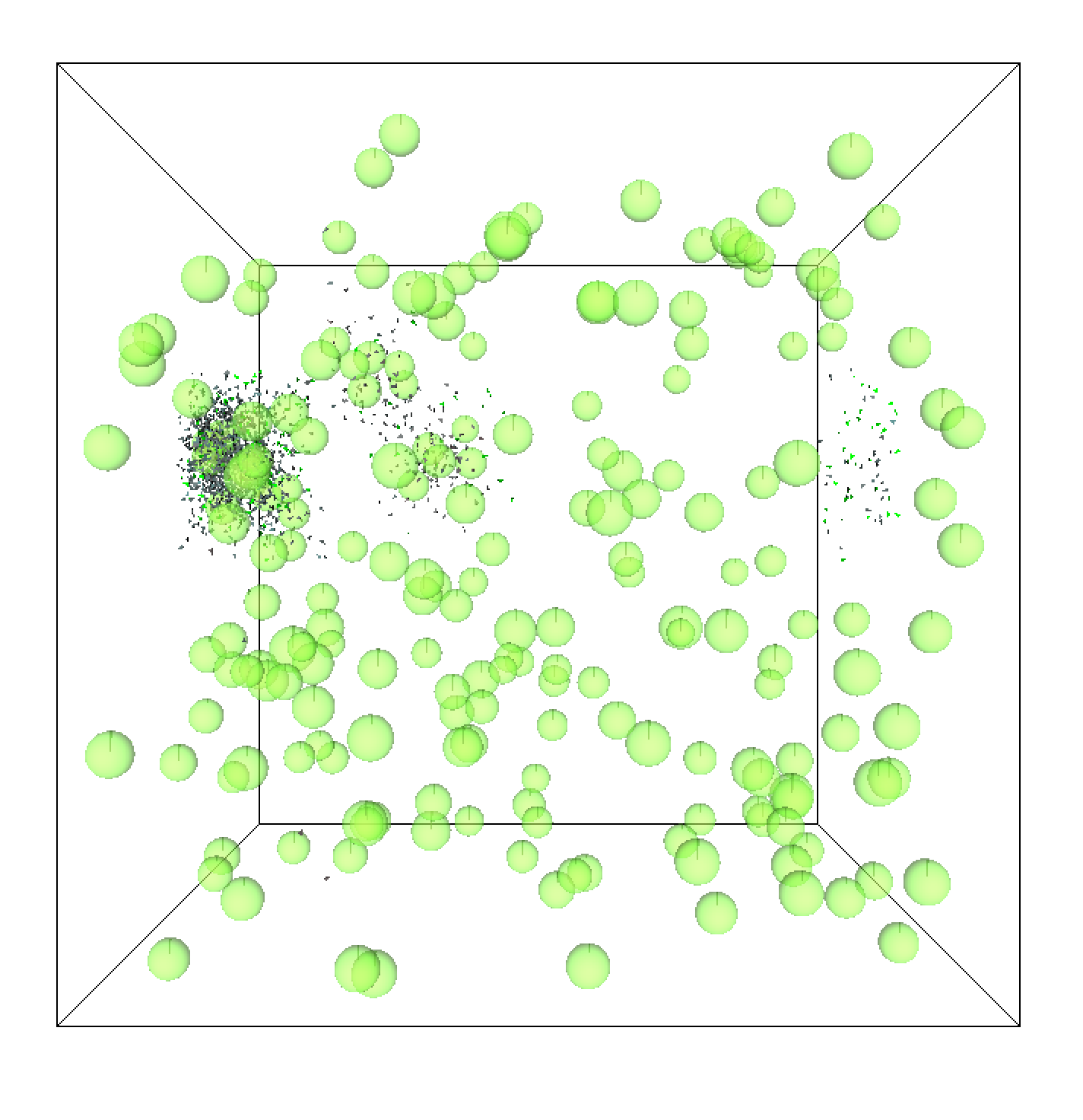}
\caption{Graphical representation of the world in a neural boids simulation. Each agent is represented as an arrow indicating its current direction. The color of an agent indicates the average value of its internal nodes. The green spheres represent the centers for energy sources. Although variants presented later in the paper display slightly different graphics, the backbone is the same.
}
\label{fig:worldsim}
\end{center}
\end{figure}

In our model, like in Reynolds' model, the population of agents moves around in a continuous three-dimensional space, with periodic boundary conditions (Figure \ref{fig:worldsim}). However, instead of using fixed rules to control the boids' motion, we allow agents to evolve their own controllers through concurrent evolutionary computation. Each agent, instead of responding to simple rules, is controlled by its own neural network. The parameters of the neural network are encoded in a genotype vector, which determines the individual's sensorimotor choices at each moment in time. This corresponds to standard evolutionary robotics methodologies \citep{nolfi2000evolutionary}, although we introduce the following variant. The genotype is evolved through the course of the simulation, via a continuous variant of an evolutionary algorithm \citep{witkowski2014a}, that is, agents with high level of fitness are allowed at any point to replicate with mutation in the middle of the running simulation.

This model also builds up on prior work on the effect of self-organized inter-agent communication and cooperative behavior on the performance of agents to solve tasks \citep{prokopenko2006evolving, olson2013predator}. Previous research has shown the difficulty of using communication channels \citep{rasmusen1994games, mitri2013using} but showed cooperative value of information transfers \citep{witkowski2016emergence}. This will be complemented by the results from previous information-theoretic analyses of learning systems, which managed to shed light on the role of information flows in learning \citep{tishby2015deep, lin2017does, tishby2011information}.

\begin{figure}[ht!]
\begin{center}
\includegraphics[width=.7\textwidth]{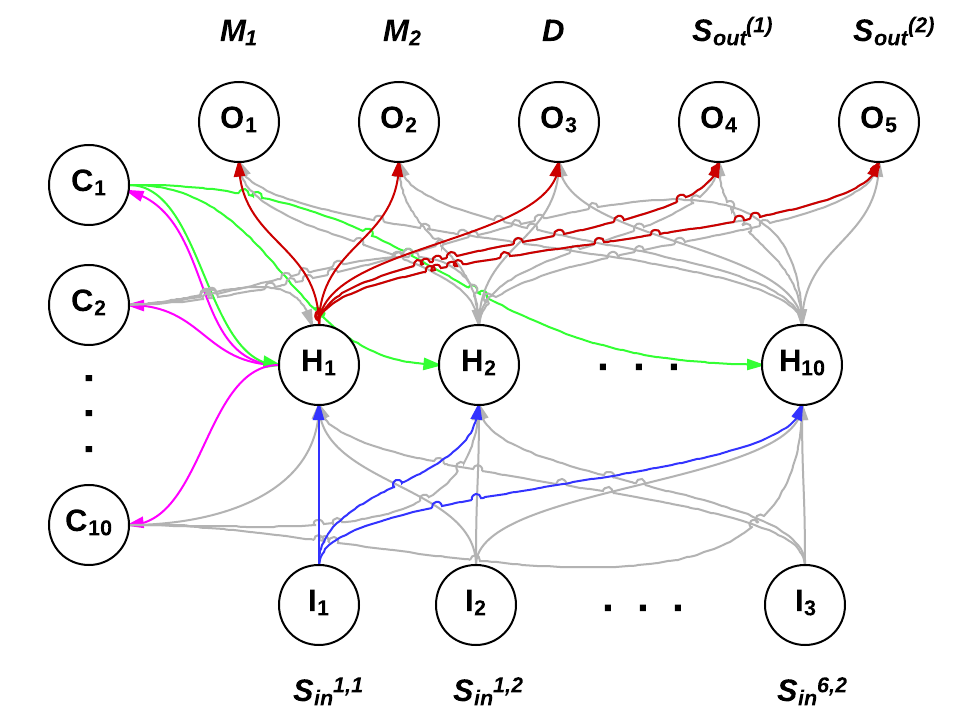} 
\caption{
{\bf Architecture of the agent's controller.} The network is composed of 12 input neurons, 10 hidden neurons, 10 context neurons and 5 output neurons.}
\label{fig:ann}
\end{center}
\end{figure}

Agents are given a certain energy, that also acts as their fitness. This will be specific to the study cases. Each agent comes with a set of 12 different sensors. The neural network (represented on Figure \ref{fig:ann}) takes the information from those sensors as inputs, in order to decide the agent's actions at every time step. The possible actions amount to the agent's motion, and in the specific variant shown here, a Prisoner's Dilemma action (cooperate or defect), as well as two output signals. The architecture is composed of a 12 input, 10 hidden, 5 output, and 10 context neurons connected to the hidden layer (see Figure \ref{fig:ann}). 

The agents' motion is controlled by $M_1$ and $M_2$, outputting two Euler rotation angles: $\psi$ for pitch (i.e. elevation) and $\theta$ for yaw (i.e. heading), with floating point values between $0$ and $\pi$. Even though the agents' speed is fixed, the rotation angles still allow the agent to control its average speed (for example, if $\psi$ is constant and $theta$ equals zero, the agents will continuously loop on a circular trajectory, which results in an almost-zero average speed over 100 steps). 

The outputs $S_\text{out}^{(1)}$ and $S_\text{out}^{(2)}$ control the signals emitted on two distinct channels, which are propagated through the environment to the agents within a neighboring radius set to $50$. The choice for two channels was made to allow for signals of higher complexity, and possibly more interesting dynamics than greenbeard studies \citep{gardner2010}.

The received signals are summed separately for each direction (front, back, right, left, up, down), and weighted by the squared inverse of the emitters distance. This way, agents further away have much less impact on the sensors than closer ones do. Every agent is able to receive signals on the two emission channels, from 6 different directions, totalling $12$ different values sensed per time step. For example, the input $S_\text{in}^{(6,1)}$ corresponds to the signals reaching the agent from the neighbors below.

The evolution is performed continuously over the population. Agents with negative or zero energy are removed, while agents with energy above a threshold are forced to reproduce, within the limits of one infant per time step. The reproduction cost is low enough, considering the threshold, to not put the life of the agent at risk.

\section{Study cases}

We go over the application of this model in three selected examples of studies. Each of them highlights a specific property for OEE. The first model shows how agents can form patterns to accelerate their search for energy, distributed over an n-dimensional space, collaborating via local signaling with their neighbors. The second study shows the invention of dynamical group strategies in a spatial prisoner's dilemma, allowing for specific cooperation effects. The third example shows the impact of growth on the emergence of noise-canceling effects.

\subsection{OEE via collective search based on communication}

Since Reynold's boids, coordinated motion has often been reproduced in number of artificial models, but the conditions leading to its emergence are still subject to research, with candidates ranging from obstacle avoidance to virtual leaders. The relation of spatial coordination and group cooperation has long been studied in game theory and evolutionary biology.

We here apply our model of agents exchanging signals and moving in a three-dimensional environment, to a task of dynamical search for free energy in space \citep{witkowski2014a, witkowski2016emergence}. Each agent's movements are controlled by artificial networks, evolved through generations of an asynchronous selection algorithm. At the term of the evolution, the agents are able to communicate to produce cooperative, coordinated behavior.

Individuals develop swarming using only their ability to listen to each other's signals. The agents are selected based on their performance at finding invisible resources in space giving them fitness. 
The agents are shown to use the information exchanged between them via signaling to form temporary leader-follower relations allowing them to flock together. The swarmers outperform the non-swarmers at finding the resource, thus reaching a neutral evolutionary space which leads to a genetic drift.

This work constructs an adaptive system to evolve swarming based only on individual sensory information and local communication with close neighbors. This addresses directly the problem of group coordination without central control, without being aware of the position direct neighbors, nor any use of the substrate where to deposit information (stigmergy) \citep{hauert2009evolved}. 
The approach has also the advantage of yielding original and efficient swarming strategies. A detailed behavioral analysis is then
performed on the fittest swarm to gain insight as to the
behavior of the individual agents.

The results show that agents progressively evolve the ability to flock through communication to perform a foraging task. We observe a dynamical swarming behavior, including coupling/decoupling phases between agents, allowed by the only interaction at their disposal, that is signaling. Eventually, agents come to react to their neighbors' signals, which is the only information they can use to improve their foraging. This can lead them to either head towards or move away from each other. While moving away from each other has no special effect, moving towards each other, on the contrary, leads to swarming. Flocking with each other may lead agents to slow down their pace, which for some of them may keep them closer to a food resource. This creates a beneficial feedback loop, since the fitness brought to the agents will allow them to reproduce faster, and eventually multiply this type of behavior within the total population.

\begin{figure}[ht!]
\begin{center}
\includegraphics[width=.9\textwidth]{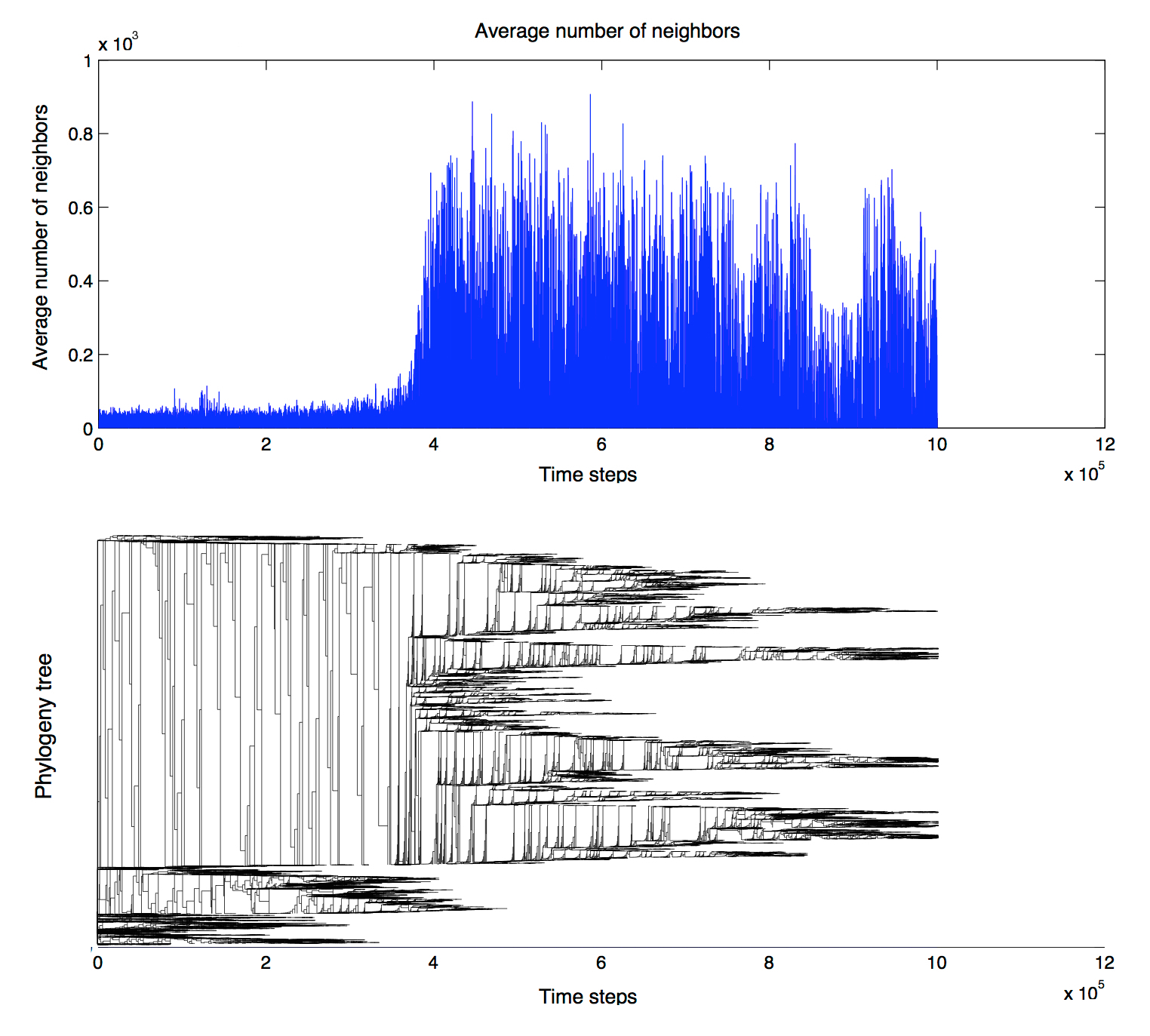}
\caption{\emph{Top: }average number of neighbors during a single run. \emph{Bottom: }agents phylogeny for the same run. The roots are on the left, and each bifurcation represents a newborn agent.}
\label{fig:treehorizontal}
\end{center}
\end{figure}

The algorithm converges to build a heterogeneous population, as shown on Figure \ref{fig:treehorizontal}. The phylogeny is represented horizontally in order to compare it to the average number of neighbors throughout the simulation. The neighborhood becomes denser around iteration $400k$, showing a higher portion of swarming agents. This leads to a firstly strong selection of the agents able to swarm together over the other individuals, a selection that is soon relaxed due to the signaling pattern being largely spread, resulting in a heterogeneous population, as we can see on the upper plot, with numerous branches towards the end of the simulation.

In this scenario, agents do not need extremely complex learning to swarm and eventually get more easily to the resource, but rather rely on dynamics emerging from their communication system to create inertia and remain close to goal areas.

\begin{figure}[ht!]
\begin{center}
\includegraphics[width=.85\textwidth]{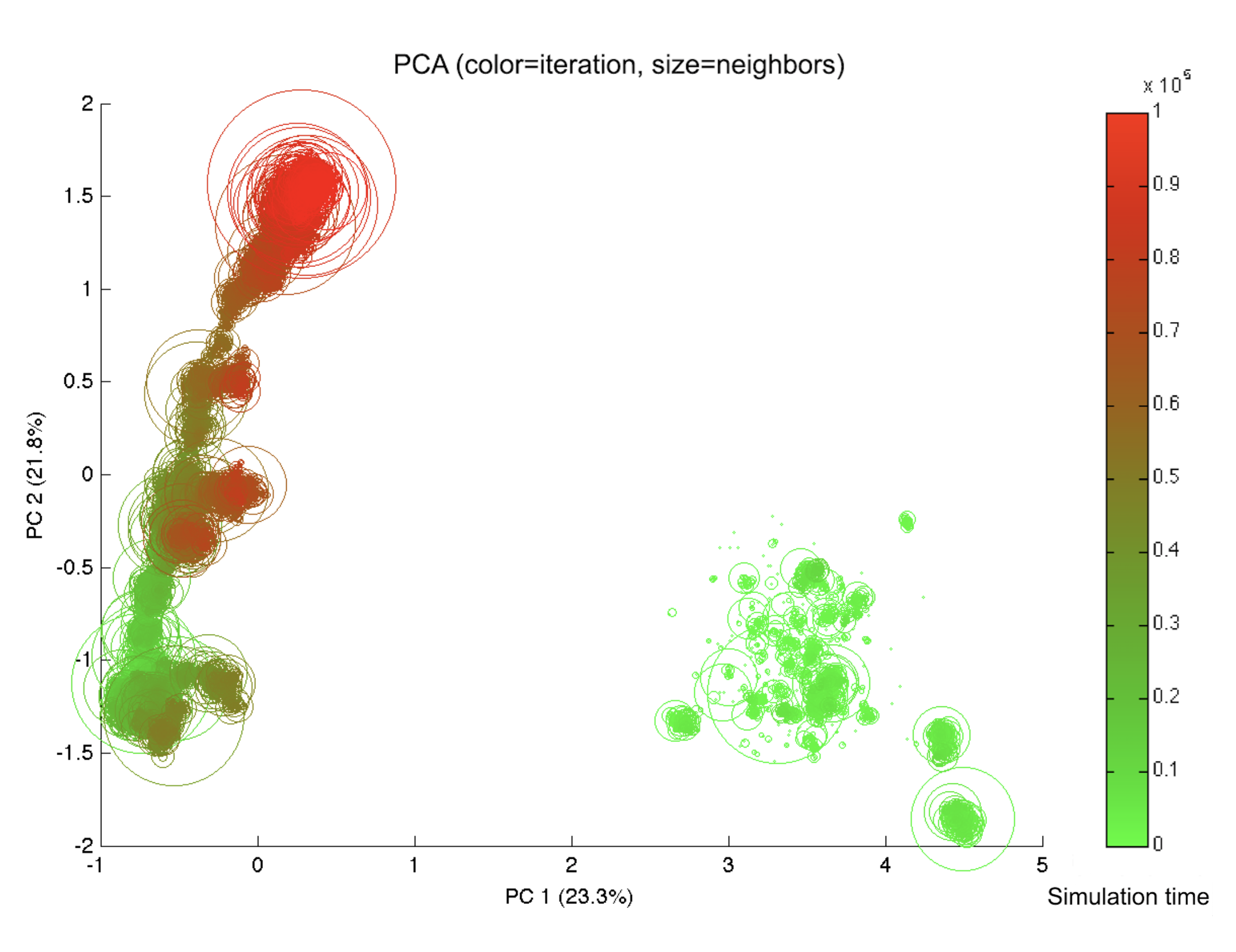}
\caption{
Two principal components of a PCA on the genotypes of all agents of a typical run, over one million iterations. Each circle represents one agent's genotype, the diameter representing the average number of neighbors around the agent over its lifetime, and the color showing its time of death ranging from bright green (at time step $0$, early in the simulation) to red (at time step $10^6$, when the simulation approaches one million iterations).
}
\label{fig:pca}
\end{center}
\end{figure}

The simulated population displays strong heterogeneity due to the asynchronous reproduction schema, which can be seen in the phylogenetic tree (Figure \ref{fig:treehorizontal}). The principal component analysis plotted on Figure \ref{fig:pca} shows a large cluster (left side) in addition to a series of smaller ones (right side). The genotypes in the early stages of the simulation belong to the right clusters, but get to the left cluster later on, reaching a higher number of neighbors. The plot shows a diverse set of late clusters, which translates to numerous distinct behaviors in the late stage of the simulation.

Such heterogeneity may suppress swarming but the evolved signaling helps the population to form and keep swarming. The simulations do not exhibit strong selection pressures to adopt specific behavior apart from the use of the signaling. Without high homogeneity in the population, the signaling alone allows for interaction dynamics sufficient to form swarms, which proves in turn to be beneficial to get extra fitness.

These results represent an improvement on previous models using hard-coded rules to simulate swarming behavior, as they are evolved from very simple conditions. Our model also does not rely on any explicit information from leaders, like previously used in part of the literature \citep{cucker2008flocking, su2009flocking}. It does not impose any explicit leader-follower relationship beforehand, letting simply the leader-follower dynamics emerge and self-organize. In spite of being theoretical, the swarming model presented in this paper offers a simple, general approach to the emergence of swarming behavior once approached via the boids rules. 
This simulation improves on previous work because agents naturally switch leadership and followership by exchanging information over a very limited channel of communication.
Finally, our results also show the advantage of swarming for resource finding. It's only through swarming, enabled by signaling behavior, that agents are able to reach and remain around the goal areas.

In terms of cooperation, this model exemplifies a case of multilevel selection theory \citep{wilson1994reintroducing, traulsen2006evolution}, which models the layers of competition and evolution, within an ecological system. Our system shows the emergence of different levels which function cohesively to maximize reproductive success. The fitness value of the group level dynamics outweighs the competitive costs, resulting in individuals constantly innovating in ways they are cooperating in a non-trivial way, to create behaviors which are not centrally coded for.

This study shows swarming dynamics emerging from a communication system between agents, immersed in a simulated environment with spatial distribution of energy resource. The concurrent evolution scheme, running at the same time as the simulation itself, led to decentralized leader-follower interactions, which allowed for collective motion patterns, which in turn significicantly improved the groups' fitnesses.

This model encodes the stochastic evolution of a controller that maps sensory input onto motor output, to optimize the performance on a task, as framed broadly by Nolfi and Floreano (2000) \cite{nolfi2000evolutionary}. We capture the fight against a difficult wall \citep{schmickl2016sooner}, which simulations typically fail at because they suddenly hit a so-called ``wall of complexity'': trivial tasks are solved easily, but it's hard to jump to solving difficult ones. If we take the no-free-lunch argument from Wolpert and Macready (1997) \cite{wolpert1997no} that no optimisation algorithm is at all times superior to others, it is natural that the more specific the algorithm, the more it is likely to fail with new problems.

Our results suggest that novelty can be produced by the asynchronous evolution of a heterogeneous community of agents, which through their mixture of strategies, may achieve open-ended, uninformed learning. The heterogeneity present in the model also offers an extension to the advantages of particle swarm optimization (PSO) \citep{eberhart1995new}. While PSO only offers one unique objective function to optimize, each agent in the swarm effectively runs its own function, which are combined into a swarm behavior. Although these results suggest open-endedness, it is worth noting we do not bring a proof that the phenomenon is truly open-ended, which may require the emergence of ever-complexifying communication, or an uninterrupted sequence of evolutionary innovations.

The information flows were a focus of the original work \citep{witkowski2016emergence}. From these flows, one can notice three main bottlenecks. The evolutionary computation contains a bottleneck effect, as a result from the selection based on the agents' performance on the task. Another bottleneck can be found between the sensory inputs of each agent, and its motor outputs, as the neural controller acts as a filter for the information. The agents' signaling also naturally contains a bottleneck effect, as the information transmitted from agent to agent is constrained by the physical communication bandwidth. The combination of these three bottlenecks allows for relevant information to be filtered into the swarm, which is able learn certain behaviors (see also next section).

\subsection{OEE via cooperative flocking}

The evolution of cooperation is studied in game theory, and stretches have been made to include spatial dimensions. This problem is often tackled by using simple models, such as considering interactions to be a game of Prisoner's Dilemma (PD).

We examined a variation of the model with a distinct fitness function in a separate study, based this time on the agents playing a spatial version of the Prisoner's Dilemma \citep{witkowski2014b}. We study the impact of the movement control on optimal strategies, and show that cooperators rapidly join into static clusters, creating favorable niches for fast replications. It is also noted that, while remaining inside those clusters, cooperators still keep moving faster than defectors. The system dynamics are analyzed further to explain the stability of this behavior.


This work presents, in an even more explicit fashion than the previous study, a model aimed at showing emergent levels of selection for cooperative behavior \citep{wilson1994reintroducing}. At every time step, agents are playing a N-player version of the prisoner's dilemma with their surrounding, meaning that they make a single decision that affects all agents around them. They get reward and/or punishment based on the number of cooperator around them. Their decision is one of the outputs of their neural network. Effectively, the payoff matrix we used is an extension of Chiong and Kirley (2012) \citep{chiong2012}, where we added the distance to take into account the spatial continuity.

Based on the outcome of the match, agents can choose a new direction, which is similar to leaving the group in the walk away strategy \citep{aktipis2004}, the main difference being that, in our case, it is also possible for groups to split. It is also similar in another aspect: there is a cost to leaving a group, as a lone agent may need time to meet others.

\begin{figure}[ht!]
\begin{center}
\includegraphics[width = 0.75\textwidth]{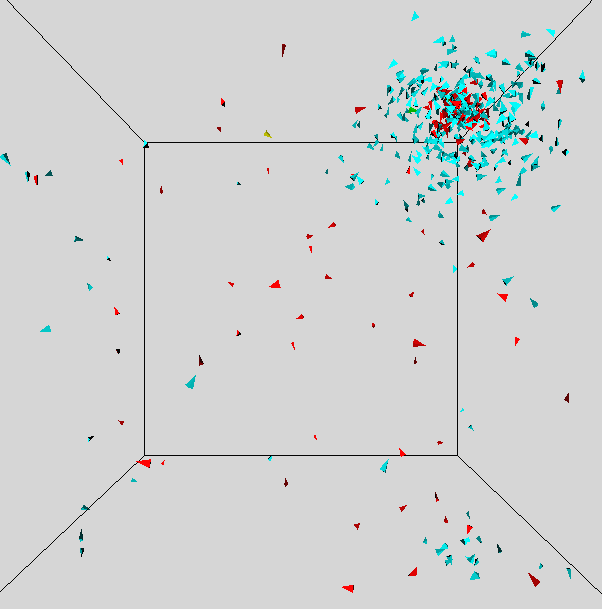}
\caption{Graphical representation of the world in a simulation. Each agent is represented as an arrow indicating its current direction. The color of an agent indicates its current action, either cooperation (blue) or defection (red). Note the cluster of cooperators being invaded by defectors.}
\label{fig:world}
\end{center}
\end{figure}

At the beginning of each run, the environment is seeded with random agents. Since all weights in their neural network are set at random, roughly half of the agents initially choose to cooperate while the other half choose to defect. This leads to a fast extinction of cooperators, until approximately 50000 time steps), until a group emerges strong enough to survive. 
The second phase follows, in which cooperators are quickly increasing in number due to the autocatalytic nature of this strategy. 
A third step happens eventually, where defectors invade the cluster, followed either by the survival of the cluster due to cooperators running away or a reboot of the cycle. In case of survival, oscillations in the proportion of cooperators can be observed. However, this phenomenon is averaged away over multiple runs, since period and phase of the oscillations are not correlated from one experiment to the other. 
Were a defector to appear near a cluster of cooperators, the cluster would react by ``reproducing away''. However, the chances to be overtaken by the defectors is much higher than in the dynamic case.

From this three-dimensional model of agents playing the Prisoner's Dilemma, the first result is that cooperators, when they are present, quickly evolve to form clusters as they represent a favorable pattern. The clustering behavior can be interpreted as a degenerated version of the simulations presented above, since the cooperating agents present the same capacities of information exchange as that model. We note that this solution is evolved through a longer time scale, as it is not always viable locally, depending on the distribution and behavioral thresholds of defectors. 
While the clustering itself can be expected, it is interesting to observe that their overall movement rate is still higher than defectors. This is even more surprising considering that those clusters do not seem to move fast. Instead, analysis shows that cooperators are moving quickly inside the cluster, which may be a way to adapt to an aggressive environment. 

In addition, comparison with the static case showed that movement made the emergence of cooperators harder, but more stable in the long run. Since it is harder for defectors to overtake a cluster of cooperators, our systems often show a soft bistability, meaning that they will eventually switch from one state to the other. It is even possible to observe a sort of symbiosis, where cooperators are generating more energy than necessary, which is in turn used by peripheral defectors. In this case, replacement rates allow cooperators to stay ahead, keeping this small ecosystem stable. 
This cohesion among cooperators seems to be enhanced by signaling, even though signals might attract defectors. Additional investigation on the transfer entropy, for instance, could be a promising next step.

Another result is found in the choice of actions, generated by the neural networks without consideration of the past actions. We notice the emergence of a dynamical memory effect, that otherwise requires to be encoded in each agent, here emerging from the agents' motion in space. 

Since the Prisoner's Dilemma game has become a common model used in evolutionary biology to study the outcomes depending on the costs that characterize an ecosystem, this model, with a fitness based on the results of such game, showed the emergence of spatial coordination based on a the exchange of signals between agents. The signals remained very simple, and the environment was fixed in time.

This model's evolutionary computation reached solutions composed of different parts, including soft bistable strategies, different radiuses of clusters, as well as the use of dynamical patterns to improve their fitness. The solutions were also distributed over different timescales. The communication between agents also allowed them to converge on these behaviors more quickly. These elements refer back to our 3-C arguments earlier, for the discovery of novel solutions to a simple PD game.

Lastly, we note that many different neural architectures may coexist as only a part of the neural architecture is used to implement flocking. This neural heterogeneity is something we’d like to insist in the context of OEE. Additionally, communication is important to filter out the neural architecture heterogeneity, which potentially holds the heterogeneity in a community (i.e. agents can stay in a community if they can communicate to each other). The communication may therefore indirectly help preserving the heterogeneity.

\subsection{OEE via large scale swarms}

Studying flocking models can also lead to the emergence of OEE, by focusing on emergent phenomena as macroscopic layers of patterns and structures that appear as a result of cooperative phenomena between autonomously behaving elements. A group of elements creates a self-organizing structure, which governs the individual micro rules and creates a new macro structure. Therefore, consecutive micro–macro recurrent self-organization is defined as an emergent phenomenon.

Here, we describe the contribution of the same swarming simulation, scaled up, showcasing the effect of size on emergence of open-endedness \citep{drozd2016critical}. Before that, we start by presenting a degenerated version of that model, which shows large-scale dynamics in the less computationally costly case of agents that don't preserve any internal state other than position and velocity \citep{ikegami2017life}.

\begin{figure}[ht!]
\begin{center}
\includegraphics[width=.7\columnwidth]{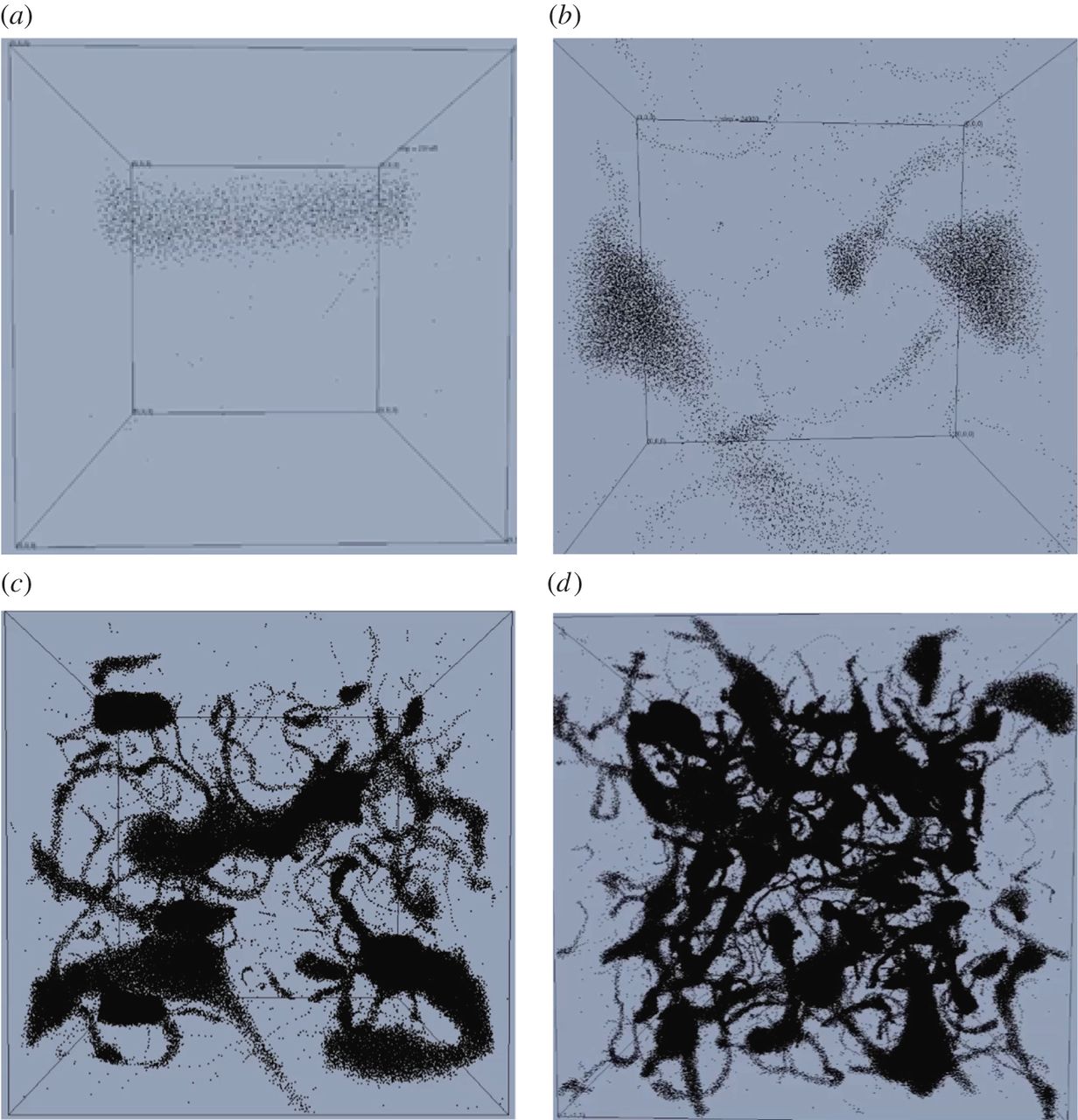}
\caption{Visualization of swarming behavior, simulated by a large scale stateless boids model \citep{ikegami2017life}. The total number of boids in each panel is (a) 2048, (b) 16 384, (c) 131 072 and (d) 524 288, respectively. Some flocks are composed of a very large number of boids with narrow filament patterns. 
The initial velocity of each boid is set at random, and the density of the total number of boids is kept constant at 16,384 (number per cubic unit). The minimum and the maximum speed are set at 0.001 and 0.005 (unit per step), respectively.
}
\label{fig:largeboids}
\end{center}
\end{figure}

Starting with this simpler stateless model, we observe a noticeable change when the total number of boids increases from 2048 to 524,288, while the density is kept constant (Figure \ref{fig:largeboids}). In order to compute large swarming behaviour, we parallelized the computational steps using the general-purpose computing on graphics processing units (GPGPU) method. The next step was to extend it to stateful agents.

We explore the effect of reaching a critical mass, and how it impact the efficiency of the swarm's foraging behavior. In particular, we study the problem of maintaining the swarm's resilience to noisy signals from the environment. To do so, we look at stateful boids, i.e. moving agents controlled by neural network controllers, which we evolve through time in order to explore further the emergence of swarming, like in the previous two sections. However, we now ground our model in a more realistic setting where information about the resource location made partly accessible to the agents, but only through a highly noisy channel. The swarming is shown to critically improve the efficiency of group foraging, by allowing agents to reach resource areas much more easily by correct individual mistakes in group dynamics. As high levels of noise may make the emergence of collective behavior depend on a critical mass of agents, it is crucial to reach in simulation sufficient computing power to allow for the evolution of the whole set of dynamics.

Because this type of simulations based on neural controllers and information exchanges between agents is computationally intensive, it is critical to optimize the implementation in order to be able to analyze critical masses of individuals. In this work, we address implementation challenges, by showing how to apply techniques from astrophysics known as treecodes to compute the signal propagation, and efficiently parallelize for multi-core architectures.
The results confirm that signal-driven swarming improves foraging performance. The agents overcome their noisy individual channels by forming dynamic swarms. The measured fitness is found to depend on the population size, which suggests that large scale swarms may behave qualitatively differently.

The minimalist study presented in this paper together with crucial computational optimizations opens the way to future research on the emergence of signal-based swarming as an efficient collective strategy for uninformed search. Future work will focus on further information analysis of the swarming phenomenon and how swarm sizes can affect foraging efficiency.

In this model, we specifically focus on the addition of noise to the food detection sense that the agents possess, and hypothesize that it can be overcome by the emergence of a collective behavior involving sufficiently large groups of agents.

Many systems, from atomic piles to swarms, seem to work towards preserving a precarious balance right at their critical point \citep{bak2013nature}. An atomic pile is said to be ``critical'' when a chain reaction of nuclear fission becomes self-sustaining. A minimal amount of fissionable material has to be compacted together to keep the dynamics from fading away. The notion of critical mass as a crucial factor in collective behavior has been studied in various areas of application \citep{marwell1993critical, oliver2001whatever}. 

Similarly, the size of the formed groups of agents may be crucial, in order to reach a critical mass in swarms, enough to overcome very noisy environments. Part of the focus will therefore be on the optimization of the computer simulation itself, as large-scale swarms may qualitatively differ in behavior from regular-sized ones.

The model extends the original setup described before, which proposed an asynchronous simulation evolving a swarming behavior based on signaling between individuals
. However, unlike the original model, where the individuals don't perceive directly either the food patches or the other agents around them, here we give a sense of vision to every agent, allowing them to detect nearby resources. However, we add a high level of noise to make this information highly imperfect.

We used an agent-based simulation to show how signal-driven swarming, emerging in an evolutionary simulation such as in Witkowski and Ikegami (2014) \cite{witkowski2014a}, allow agents to overcome noisy information channels an improve their performance in a resource finding task. Our first contribution is the very introduction of noise, demonstrating that the algorithm performs well against noises filling up channels of information almost up to their full capacity, in the inputs of agents. The individuals, by means of a swarming behavior helped by basic signaling, manage to globally filter out the noise present in the information from their sensory inputs, to reach the food sites.

We proposed a hierarchical method based on the Barnes-Hut simulation in computational physics and its parallel implementation. We achieved a performance improvement of a few orders of magnitude over the previous implementation \citep{witkowski2014a}. This implementation is crucial to achieve the simulation of a sufficient number of agents to test for large-scale swarms (i.e. involving a very large number of individuals), which have been suggested to generate qualitatively different dynamics.

The optimization of the fitness acquired by phenotypes using efficient patterns of behavior (motion and signaling), which themselves are encoded in the weights of agents' neural networks. The real optimization therefore occurs at the higher level of the darwinian-like process in the genotypic search space. Efficient genotypes are selected by the asynchronous genetic algorithm throughout a simulation run.

\begin{figure}[ht!]
\begin{center}
\includegraphics[width=.7\columnwidth]{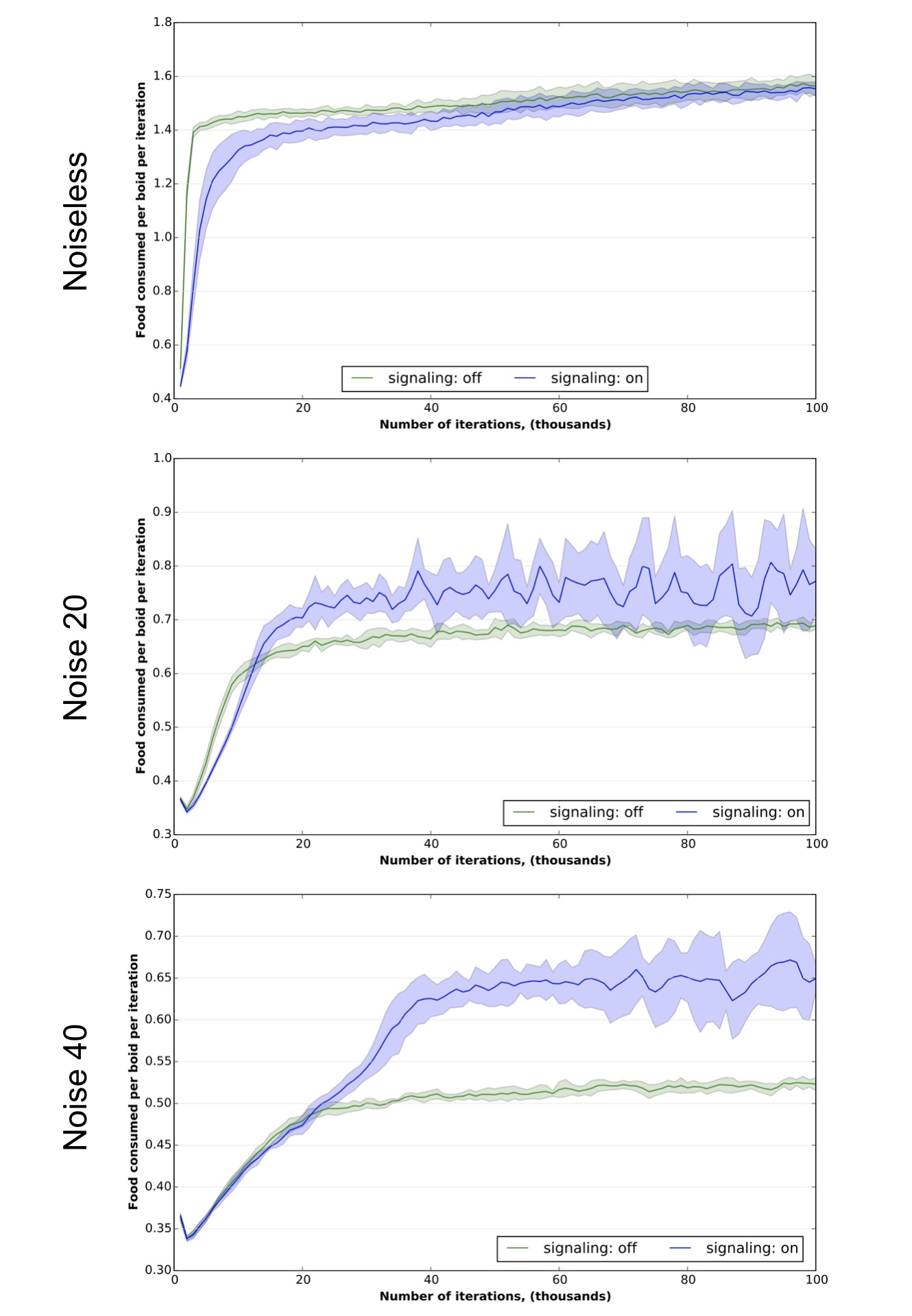}
\caption{Agents' efficiency plot with and without signal, from a stateful (neural-network-controlled) boids model in the original work \citep{drozd2016critical}, with mean (central line) and standard deviation range (area plot) over 10 runs. The plots correspond to noiseless (top), constant noise 20 (middle), and constant noise 40 (bottom), respectively.}
\label{fig:boidsefficiency}
\end{center}
\end{figure}

We observed that signaling improves the foraging of agents (see Figure \ref{fig:boidsefficiency} for plots from Drozd et al. \citep{drozd2016critical} of efficiency or fitness against simulation time), the average resource retrieved per agent per iteration as a measure of the population's fitness. Without noise, the agents using signaling are less efficient than their silent counterpart, which we found is not due to the cost of signaling, but rather because of the excess of noise brought by the signal inputs. The difference remains very small between signaling and non-signaling agents.

We find however that from a certain noise level, the cost to signal is fully compensated by the benefits of signaling, as it helps the foraging of agents. The average fitness becomes even higher as we increase the noise level, which suggest that the signaling behavior increases in efficiency for high levels of noise, allowing the agents to overcome imperfect information by forming swarms.

We also observe scale effects in the influence of the signal propagation on the average fitness of the population. For a smaller population, only middle values of signal propagation seem to bring about fitter behaviors, whereas this is not the case for larger sizes of population. On the contrary, larger populations are most efficient for lower levels of signal propagation. This may suggest a phase transition in the agents' behavior for large populations, eventually in the way the swarming itself helps foraging.

Understanding criticality seems strongly related to a broad, fundamental theory for the physics of life as it could be, which still lacks a clear description of how it can arise and maintain itself in complex systems. The effects of criticality have recently been investigated futher by one of the authors, using a similar setup \citep{khajehabdollahi2018critical}. The results showed exploratory dynamics at criticality in the evolution of foraging swarms, and the tension between local and global scale adaptation.

Through this work, by increasing the number of simulated boids that maintain their own states, we may introduce more than the mere number. By allowing for many information exchanges between computing agents, the simulation can effectively take leaps of creativity. In Stanley and Lehman's 2015 book \cite{stanley2015greatness}, objective functions are presented as a distraction, as novelty and diversity might not be achieved by hard-coding the arrival point. Here, in contrast, we have many evolvable objective functions cooperate in reaching a solution, as a stepping stone to reach the search for novelty.

By letting the swarm grow, we see the emergence of collective intelligence, which corresponds to the invention of signal-based error correction. By exchanging signals, the agents are able to correct the error induced by the noise we injected in the simulation. Like for the large scale boids simulation, the invention happens after a critical mass of agents is reached, suggesting similar dynamics with stateful agents.

\section{Discussion}

OEE is the everlasting innovative processes found in human technologies and biological evolutions, and we barely observe open-endedness outside these examples. Yet, some artificial systems demonstrate close-to-OEE phenomena, which we have discussed earlier in this paper.

Achieving real OEE remains an open challenge, and at this point all works in the literature works fall short of that objective. Although that may be the case with the swarm models presented above, it was one of our goals to emphasize the importance of maintaining the evolvability of a system. In an adversarial game theoretical setup for example, reaching an ESS or a local attractor may keep the system from inventing new solutions. In this sense, explicitly stopping the system from learning too much may allow the system to avoid being stuck in such attractor, and possibly keep innovating forever.

In this paper, we propose collective intelligence as a driving force towards open-ended evolution, suggesting that collective groups can develop the ability to be more innovative. Instead of aiming at optimizing one fixed objective function, a collective swarm of agents works with as many competing objectives as there are agents in the swarm.

Through information exchanges between a certain number of agents, these objectives, embodied in the agent's behaviors, can collaborate to implement a search for novelty. All agents contribute to the search in behavioral space, as one whole organization, by exploring the adjacent possible. 
Each novel discovery in the system, or emergent level of organization, can be reached from an adjacent state where the system was previously. The way one moves from one point to the next, which should retain information accumulated in the past, is constrained by the structure of the swarm, in a bottleneck effect.

We discuss several instances of bottlenecks in this paper. One is a task or environmental condition which each agent must overcome. In the case of foraging environment, organizing swarming turned out to be a critical step. So it became a major transition from non-swarming to swarming agents. Swarming behavior was obtained by  organizing a hill side function by the neural controller. After the swarming behavior has been achieved, other properties (e.g. individual pattern) start to evolve. So for the task, swarming was a necessary behavior to organize and was a bottleneck for the entire evolutionary process. In other words, OEE emerges by setting up a right environmental condition.

In case of a game-theoretic situation, such as Study 2, the communication system among all agents constituted a bottleneck to achieve mutual cooperation. With the emergence of niche construction, the door opens for regulating mechanisms such as cooperation, reciprocal altruism or social punishment, to get implemented. In this example, the OEE in terms of the invention of cooperation mechanisms, can only evolve as a secondary structure once the swarming structure is already established.

In the case of large swarm models, the bottleneck is twofold. One is scale itself and also its CPU resource. We discussed the evolution of swarm by increasing its size, showing there is a critical size where the different kinds of fluctuation dominates in larger flocks (i.e. heading direction fluctuation to the density fluctuation). If such a transition occurs at the larger size simulation (we expect it can happen in each 3-4 order difference in sizes), we say that OEE is caused by increasing the size.

In addition to this point, 3D swarm models require a huge computational power and we need to elaborate programming for a large scale systems. In Study 3, each boid has a list of neighbors and it is updated periodically. This speeds up the calculation of the distance from the one to its neighboring boids. In study 1, each bird can listen to the sound sent unidirectionally from the other boids, so that we don’t have to calculate the exact distance. Real birds will never measure the distance to the other individuals. So measuring the distance is an unnecessary bottleneck due to the computational model. Here the OEE is the new computational techniques to overcome this computational bottlenecks.

The computation of a swarm displays a bottleneck effect, in the sense that the emergent properties of the swarm and its embodiment in a simulated environment may constrain the way the information (communication, lineage information) flows within the system, and the way relevant information (strategies, motion patterns) is progressively retained\footnote{A swarm can be shown to act as a collective memory, either explicitly/statefully \citep{witkowski2016emergence} or dynamically/statelessly \citep{couzin2002collective}.} through time in its structure. Nevertheless, the simplicity of such information flows may be limiting, more complex information transfer protocols may need to emerge from bottlenecks to bootstrap OEE.

For open-endedness, bottlenecks are crucial to, perhaps counterintuitively, act against learning. We observe examples of such bottlenecks in systems like Picbreeder \citep{secretan2008picbreeder}, where one must find a way to avoid the system from assuming that the current apparent goal is the ultimate goal, as this would preclude further innovations. Picbreeder-like systems present similarities with our signal-based swarms, as they have communication between many agents filter information to let innovations come about.

As suggested in the beginning of the paper, bottlenecks can be caused by different components: an explicit communication system, a concurrent evolutionary system, and a greater complexity. These three components are highlighted in the studies described above, and we propose them as the principal ones to create novelty and heterogeneity in solutions.

First, the communication between agents is shown to catalyze swarming and cooperation strategies. 
In previous work of turn-taking interaction between two agents installed with neural networks \citep{iizuka2003adaptive}, we noticed that performing democratic turn-taking offers novel styles of motion evolutionarily. Accordingly, here, the local interactions between agents in a flock allow for the swarm to take particular shapes (Study 1), invent an explicit cooperative protocol (Study 2), and implement a noise-canceling policy (Study 3). To reach OEE, perhaps more than mere signaling, higher complexity levels of language may need to emerge.

Second, the concurrent evolution algorithm essentially selects for meaningful information in behavioral space, by squeezing noisy behavior through a selective bottleneck. However, instead of using one unique objective function, the selection is distributed asynchronously in space and time. Differential timescales also helps accelerating the learning, which should happen as fast as possible, while retaining the way to generate the best found patterns discovered in the past. Lastly, once past the selection bottleneck, heterogeneity seems to increase considerably in genotypic space.

Third, in terms of complexity, given the population size is large enough, with a consequently large number of degrees of freedom, we notice the swarm dynamics significantly change, in various ways. The flock’s surface curvature may vary for large or small flocks, as well as the attraction and repulsion induced by the exchange of different signals. The motor responses may be amplified, since the input signals may significantly increase, given a higher density of neighboring agents, as seen in Study 1. Similarly, smaller flocks may display a more ordered behavior, with the trade-off however of being more sensitive to noise, since the critical mass is not reached to implement noise-canceling effects, as demonstrated in Study 3. Larger flocks can also be a source of individual behavioral differentiation, when a higher order of organization emerges. The key is not the size nor the amount of new information, but rather the system promoting the invention of new coordination patterns within itself.

We have shown how collective intelligence has the ability to augment the creation of new and diverse solutions in a swarm, when given limited channels of communication, a concurrent evolution bottleneck and a large number of constrainted degrees of freedom. It come as an inspiration for scientists: a good way to build an open-ended system, able to indefinitely discover new inventions, seems not to reside in centralized computation, but rather in distributed systems, composed of large collectives of communicating agents.

\section{Acknowledgements}

The authors would like to thank their collaborators who contributed partially to this work: Nathanael Aubert-Kato, Aleksandr Drozd, Yasuhiro Hashimoto, Norihiro Maruyama, Yoh-ichi Mototake and Mizuki Oka.


\bibliographystyle{apalike}
\bibliography{witkowski}









\end{document}